\documentclass[twocolumn]{svjour3}
\usepackage{amsbsy,amssymb,amsmath,bm}
\usepackage{graphicx}

\RequirePackage{fix-cm}
\smartqed
\journalname{Journal of Superconductivity and Novel Magnetism}

\begin{document}

\title{Overlapping hot spots and charge modulation in cuprates.}
\author{Pavel A. Volkov \and Konstantin B. Efetov}
\institute{Pavel A. Volkov \at
              Theoretische Physik III, Ruhr-Universit\"{a}t Bochum, D-44780 Bochum, Germany \\
              Tel.: +49-234-321-37-36\\
              Fax: +49-234-321-44-48\\
              \email{pvolkov@tp3.rub.de}                      \and
           Konstantin B. Efetov \at
              Theoretische Physik III, Ruhr-Universit\"{a}t Bochum, D-44780 Bochum, Germany\\and National University of Science and Technology ``MISiS'', Moscow, 119049, Russia}
\date{Received: date / Accepted: date}
\maketitle

\begin{abstract}
Particle-hole instabilities are studied within a two dimensional model of
fermions interacting with antiferromagnetic spin fluctuations (spin-fermion
model). In contrast to previous works, we assume that neighboring hot spots
overlap due to a shallow dispersion of the electron spectrum in the
antinodal region and include in the consideration effects of a remnant low
energy and momentum Coulomb interaction. It turns out that this modification
of the model drastically changes the behavior of the system. The leading
particle-hole instability at not very weak fermion-fermion interaction is no
longer a charge density wave with a modulation along the diagonals of the
Brillouin zone predicted previously but a Pomeranchuk-type deformation of the
Fermi surface breaking the C$_{4}$ symmetry of the system. This order does
not prevent from further phase transitions at lower temperatures. We show
that, depending on parameters of the interaction, either d-wave
superconductivity or charge density wave with modulations along the bonds of
the $CuO$ lattice is possible. The low momentum remnant Coulomb interaction
enhances the d-form factor of the charge density wave.
Comparison with experimental data allows us to conclude that in many cuprate
compounds the conditions for the proposed scenario are indeed fulfilled. Our
results may explain important features of the charge modulations observed
recently.

\end{abstract}






\section{Introduction}

\label{intro}

Recent discovery of charge density waves (CDW) in hole-doped cuprates has raised a new wave of interest to the physics of these materials. The CDWs have been detected in underdoped samples of YBCO \cite%
{y123REXS-3,y123REXS-4,y123REXS-5,y123REXS-6,y123XRD-3,y123XRD-4}%
, Bi-2201 \cite{bi2201STM-1,bi2201STM-2}, Bi-2212 \cite%
{Bi2212REXS-2,Bi2212STM-2,Bi2212STM-3}
and Hg-1201 \cite{HgREXS,HgXRD} by direct methods such as resonant X-ray
scattering, hard X-ray diffraction and scanning tunneling microscopy(STM).
A wide variety of techniques, that can be sensitive to a CDW indirectly, also confirm its presence, among them transport measurements \cite{y123Transp,HgTransp}, nuclear magnetic resonance \cite{y123NMR-2}, ultrasound propagation \cite{y123US} and pump-probe experiments \cite{y123Refl}.

Several common properties of this CDW state in hole-doped cuprates have been identified. The transition temperature $T_{CDW}$ is higher than $T_c$ but lower or equal to the pseudogap temperature $T^*$. The temperature and magnetic filed dependence of the CDW amplitude({\it e.g.} \cite{y123XRD-3}) is consistent with the CDW state competing with the superconductivity.

The CDW wave vectors seen in the experiments \cite{y123REXS-5,y123XRD-3,HgXRD,Bi2212STM-0} are directed along the $Cu-O-Cu$ bonds of the $CuO_2$ plane (axes of the Brillouin zone, axial CDW). The CDW period is approximately equal along both the axes and increases with doping \cite{y123XRD-3,bi2201STM-1,Bi2212REXS-2}.

Recent studies have also revealed important information about the distribution of the modulated charge inside the unit cell, {\it i.e.} the CDW form factor. It has been found for Bi-2212 \cite{Bi2212STM-2} and YBCO \cite{y123REXS-5} that
the charge is modulated approximately in antiphase at two oxygen sites of the unit cell with the charge at $Cu$ site being constant. In other words, the CDW form factor is characterized by a dominant d- component. The properties mentioned up to now are quite different from the stripe state of the La-based compounds \cite{StripeRev,y123REXS-6}.

Considerable attention has been also drawn to the nanoscale structure of the CDW. Quantum resistance oscillation experiments \cite{y123QO-1,y123QO-2,HgQO} have been interpreted \cite{seb1} as being due to a checkerboard modulation, where CDWs with two orientations uniformly coexist throughout the sample. Results of studies \cite{Bi2212STM-0,y123REXS-4} suggest, however, show that the charge ordered state consists of domains where CDW is unidirectional.

There have been a number of attempts to obtain the CDW state with the properties discussed above from microscopic calculations. In a model of fermions interacting with antiferromagnetic critical spin fluctuations \cite{SFREV} (spin fermion (SF) model) a charge order appears in perturbation theory as a subleading instability \cite{MetSach} hindered by the curvature of the Fermi surface. This order is a checkerboard CDW with d-form factor and wave vectors directed along the diagonals of the BZ \cite{Efetov2013,Sach2013}. The nearest-neighbor Coulomb interaction can, in principle, make this state leading as has been shown in Ref. \cite{SachSau}. Moreover, thermal fluctuations between this charge order and SC have been shown to be able to destroy both the orders while pertaining a single-particle gap \cite{Efetov2013}, which can explain the pseudogap phase. Qualitative aspects of CDW-SC competition are also well-captured in the SF model: moderate magnetic fields suppressing the superconductivity have been shown to favor CDW \cite{MEPE} resembling the experiment \cite{y123US}. The vortex cores in the SC state have been shown to contain CDW \cite{EMPE} which is seen in STM \cite{hoffman,hamidian}. The diagonal direction of modulation wavevectors contrasting the experiments, however, has proved to be quite robust.

Some proposals have been put forward to overcome this contradiction. A CDW with the correct wavevector direction has been
obtained in Refs. \cite{cascade,WangChub2014}, however the form factor has been found to lack the dominant d-symmetry with a large s- component. In Ref. \cite{pepin} a mixture of the states proposed in Ref. \cite{Efetov2013} and Ref. \cite{WangChub2014} has been considered, which should contain either the diagonal modulation or an axial CDW with a non d-form factor. CDW considerations using other models \cite{Punk2015,DavisDHLee2013,Kampf2013,yamakawa2015,Kampf2014,ChowdSach2014,ThomSach}also do not seem to explain the robustness of the axial d-form factor CDW in the cuprates.

In this contribution we review the treatment of the SF model allowing the neighboring hot spots to overlap, such that eight hot spots merge into two hot regions entirely covering the antinodal portions of the Fermi
surface. This corresponds to sufficiently small values $|\varepsilon
(\pi ,0)-E_{F}|\lesssim \Gamma $, where $E_{F}$ is the Fermi energy, $\varepsilon \left( \pi ,0\right) $ is the energy in the middle of the Brillouin zone edge, and $\Gamma $ is a characteristic energy of the
fermion-fermion interaction due to the antiferromagnetic fluctuations.
Consideration of this limit is motivated by ARPES data \cite{Bi2201ARPES-1,Bi2201ARPES-2,Bi2212ARPES-1,Bi2212ARPES-2} showing that the energy separation between the hot spots and $(\pi ,0);(0,\pi )$ is actually
quite small. In addition to the electron-electron interaction via
paramagnons, we consider also the effects of low-energy (low-momentum) part of
the Coulomb interaction, which should not contradict the philosophy of the
low energy SF model. A detailed derivation and discussion of the results can
be found in our paper \cite{preprint}.

\section{Model and main equations}
\label{sec1}

\begin{figure}[tbp]
\includegraphics[width=0.5\linewidth]{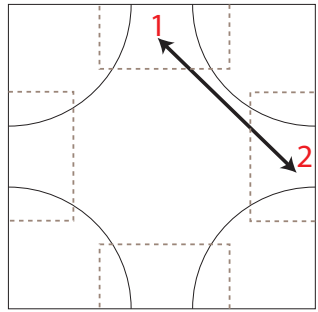}
\centering
\caption{A typical cuprate Fermi Surface with two regions connected
by the antiferromagnetic wavevector $(\pi,\pi)$ }
\label{fig1}
\end{figure}

We consider a single band of fermions interacting through critical antiferromagnetic (AF) fluctuations (paramagnons) represented by a spinful bosonic field as well as the Coulomb force. As the AF fluctuations peak at momentum transfer $(\pi,\pi)$ we restrict our model to two regions of the Fermi surface connected with this vector represented in Fig. \ref{fig1}. Inside these regions we do not specify individual hot spots, {\it i.e.} points on the FS connected by $(\pi,\pi)$ as we assume the interaction to be important in all the whole region. This assumption is supported by ARPES experiments \cite{Bi2201ARPES-1,Bi2201ARPES-2,Bi2212ARPES-1,Bi2212ARPES-2} showing that $|\varepsilon (\pi ,0)-E_{F}|$ is actually smaller than the pseudogap energy, which can be taken as the interaction scale.

The fermion-paramagnon part of the Lagrangian takes the form:
\begin{equation}
\begin{split}
L_{\mathrm{SF}} =\sum_{\mathbf{p,}\nu =1,2} \chi
_{\mathbf{p}}^{\nu \dagger } \left[\partial _{\tau }+\varepsilon _{\nu}\left(\mathbf{p}\right) -\mu_0 \right] \chi _{\mathbf{p}}^{\nu }+
\\
+\sum_{q}\vec{\varphi}_{-\mathbf{q}}(-v_{s}^{-2}\partial _{\tau }^{2}+
\mathbf{q}^{2}+\xi^{-2})\vec{\varphi}_{\mathbf{q}} \\ +\lambda
^{2}\sum_{\mathbf{p,q}}\left[ \chi _{\mathbf{p+q}}^{1\dagger }
\vec{\varphi}_{\mathbf{q}}\vec{\sigma}\chi _{\mathbf{p}}^{2}+\chi _{\mathbf{
p+q}}^{2\dagger }\vec{\varphi}_{\mathbf{q}}\vec{\sigma}\chi
_{\mathbf{p}}^{1} \right].
\end{split}
\label{sf_h}
\end{equation}
where $\varepsilon _{\nu}(\mathbf{p})$ is the electron dispersion in region $\nu=1,\;2$ (including the chemical potential), $v_s$ is the velocity of spin waves and $\xi$ is the magnetic correlation length. We shall not write explicitly the terms corresponding the Coulomb interaction as we will take their effect into account qualitatively.

Assuming that the regions 1 and 2 occupy a small portion of the BZ we expand $\varepsilon _{1(2)}(\mathbf{p})$ around $[\pi,0]([0,\pi])$ resulting in $\varepsilon _{p}^{1}=\alpha p_{x}^{2}-\beta p_{y}^{2}-\mu_0,\;\varepsilon_{p}^{2}=\alpha p_{y}^{2}-\beta p_{x}^{2}-\mu_0$, where $\mu _{0}$ is the chemical potential counted from $\varepsilon (\pi ,0)=\varepsilon (0,\pi )$. Moreover, we will average the curvature term (the one with $\beta$) inside each region leading to the final form:
\begin{equation}
\varepsilon _{p}^{1}=\alpha p_{x}^{2}-\mu ,\quad \varepsilon _{p}^{2}=\alpha
p_{y}^{2}-\mu ,  \label{mod_disp}
\end{equation}
where $\mu =\mu _{0}+\langle \beta p_{\parallel }^{2}\rangle $. To study particle-hole instabilities we define the order parameter:
\begin{equation}
W_{\mathbf{Q}}(\tau-\tau',\mathbf{k})=\langle \chi_{\mathbf{k}-\mathbf{Q}/2,\sigma}^{\dagger}(\tau')
\chi_{\mathbf{k}+\mathbf{Q}/2,\sigma }(\tau)\rangle  \label{mod_op}
\end{equation}
As has been shown in \cite{preprint} this order parameter is related to density modulations at the three atoms of the unit cell in the following way:
\begin{equation}
\begin{split}
\delta n_{Cu}(\mathbf{r}) =2e^{i\mathbf{Q}\mathbf{r}}\sum_{\mathbf{k}}W_{\mathbf{Q}}(0,\mathbf{k})+c.c,  \\
\delta n_{O_{x}}(\mathbf{r}) =\frac{p}{4}e^{i\mathbf{Q}\mathbf{r}}\sum_{\mathbf{k}}\cos(k_{x}a_{0})W_{\mathbf{Q}}(0,\mathbf{k})+c.c.,  \\
\delta n_{O_{y}}(\mathbf{r}) =\frac{p}{4}e^{i\mathbf{Q}\mathbf{r}}\sum_{%
\mathbf{k}}\cos (k_{y}a_{0})W_{\mathbf{Q}}(0,\mathbf{k})+c.c.
\end{split}
\label{mod_dens}
\end{equation}
As both the regions we consider yield approximately $\cos (k_{x}a_{0})+\cos (k_{y}a_{0})\approx 0$ we have $\delta n_{O_{x}}(\mathbf{r}) + \delta n_{O_{y}}(\mathbf{r})\approx 0$ in or model, {i.e.} charge is modulated in antiphase at the two oxygen sites of the unit cell.

Now we can discuss the qualitative effects of the Coulomb interaction in the $CuO_2$ plane. The strong on-site repulsion prohibits any real charge modulations on the $Cu$ sites leading to the constraint: $\delta n_{Cu}=0$ for the order parameter. Together with $\delta n_{O_{x}}(\mathbf{r}) + \delta n_{O_{y}}(\mathbf{r})\approx 0$ discussed above this leads to the conclusion that the charge modulations obtained in our model will have the d-form factor in accord with the experiments \cite{Bi2212STM-2,y123REXS-5}.

The nearest-neighbor Coulomb interaction has been shown in \cite{SachSau} to suppress superconductivity and support charge ordering, explaining $T_{CDW}>T_c$. This allows one to consider the particle-hole channel of the model separately from the particle-particle one.

\section{Pomeranchuk instability and intra-cell charge modulation.}
\begin{figure}[h]
\centering
\includegraphics[width=0.5\linewidth]{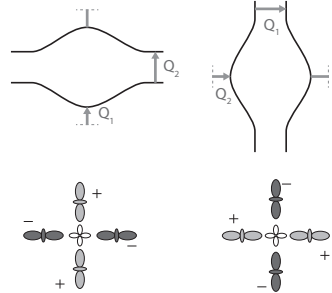}
\caption{Pictorial representation of two possible shapes of the Fermi
surface below the Pomeranchuk transition and the corresponding
intra-unit-cell charge redistributions. Grey arrows mark the emergent nesting vectors for regions $1$ and $2$.}
\label{figpom}
\end{figure}

Our main finding is that for sufficiently small $\mu$ the leading particle-hole instability is the one with ${\bf Q}=0$. The ordered state is then characterized not by a CDW, but rather a deformation of the FS (this type of transition is known as Pomeranchuk instability \cite{pomeranchuk}, \cite{yamase2005}). Moreover, it follows from (\ref{mod_dens}) that such a deformation leads to a redistribution of charge between the oxygen sites of the unit cell (see Fig.\ref{figpom}).

One can obtain this result analytically for a simplified BCS-like model where the paramagnon propagator is replaced by a constant. For that case a mean-field analysis yields that if $\mu/T_{Pom}\leq1.1$ then it is the leading instability. $T_{Pom}$ is given in this case by $\frac{1}{2\alpha }\left( \frac{\lambda _{0}\Lambda }{4\pi ^{2}}\right) ^{2}$ where $\lambda_0$ is the dimensionless coupling constant and $\Lambda$ is the size of a single region in the momentum space. This expression contrasts the usual exponential dependence obtained in BCS-like theories. A detailed account on this simplified case is presented in \cite{preprint}.

\begin{figure}[h]
\includegraphics[trim = 100 150 0 0,width=0.8\linewidth]{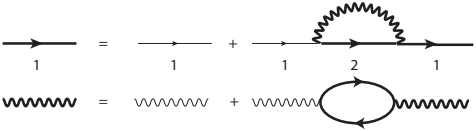}
\caption{Feynman diagrams for fermionic and bosonic propagators illustrating
the approximations used.}
\label{figFeyn}
\end{figure}

Now let us turn to the model presented here. As a starting approximation we will use the self-consistent equations represented by diagrams in Fig. \ref{figFeyn}
The integral over momentum in the fermionic self-energy can be greatly simplified provided $\mu v_{s}^{2}/\alpha \ll (v_{s}/\xi )^{2}$, i.e. that the correlation length is not too large. Then the self-energy and polarization operator do not depend on the momentum. To analyze the FS deformation we distinguish the 'even' $(\Sigma_1+\Sigma_2)/2\equiv i\varepsilon _{n}-if(\varepsilon _{n})$ and 'odd' $(\Sigma_1-\Sigma_2)/2\equiv P$ contributions to self-energy, with the latter being zero in the normal state. After the momentum integration one can introduce an energy scale $\Gamma =\left( \frac{\lambda ^{2}v_{s}}{\sqrt{\alpha }\hbar ^{2}}\right)^{2/3}$ and write the self-consistency equations in the dimensionless form (see Eq. \ref{sf_Pom}), where $\bar{a}$ denotes $\overline{(v_s/\xi)}$.
\begin{figure*}
\begin{centering}
\begin{equation}
\begin{gathered}
\bar{f}(\bar{\varepsilon}_{n})-\bar{\varepsilon}_{n}=
0.75\bar{T}\sum_{\bar{\varepsilon}_{n}^{\prime }}
\frac{1}{\sqrt{\bar{\Omega}(\bar{\varepsilon}_{n}-\bar{\varepsilon}_{n}^{\prime})+\bar{a}}}
\frac{\mathrm{sgn}(\mathrm{Re}[f(\bar{\varepsilon}_{n}^{\prime })])}{2}
\left[
\frac{1}{\sqrt{i\bar{f}(\bar{\varepsilon}_{n}^{\prime })+\bar{\mu}+\bar{P}(\bar{\varepsilon}_{n}^{\prime })}}+%
\frac{1}{\sqrt{i\bar{f}(\bar{\varepsilon}_{n}^{\prime })+\bar{\mu}-\bar{P}(\bar{%
\varepsilon}_{n}^{\prime })}}
\right]
\\
\bar{P}(\bar{\varepsilon}_{n})=i\cdot 0.75\bar{T}\sum_{\bar{\varepsilon}%
_{n}^{\prime }}\frac{1}{\sqrt{\bar{\Omega}(\bar{\varepsilon}_{n}-\bar{%
\varepsilon}_{n}^{\prime })+\bar{a}}}\frac{\mathrm{sgn}(\mathrm{Re}[f(\bar{%
\varepsilon}_{n}^{\prime })])}{2}\left[ \frac{1}{\sqrt{i\bar{f}(\bar{%
\varepsilon}_{n}^{\prime })+\bar{\mu}-\bar{P}(\bar{\varepsilon}_{n}^{\prime })}}-%
\frac{1}{\sqrt{i\bar{f}(\bar{\varepsilon}_{n}^{\prime })+\bar{\mu}+\bar{P}(\bar{%
\varepsilon}_{n}^{\prime })}}\right]
\\
\bar{\Omega}(\bar{\omega}_{n})-\bar{\omega}_{n}^{2}=-\frac{\bar{T}}{2}\sqrt{\frac{v_{s}^{2}/\alpha }{\Gamma }}
\sum_{\bar{\varepsilon}_{n}}
\left[
\frac
{\mathrm{sgn}(\mathrm{Re}[f(\varepsilon_{n})])}
{\sqrt{i\bar{f}(\bar{\varepsilon}_n)+\bar{\mu}+\bar{P}(\bar{\varepsilon}_n)}}
\frac
{\mathrm{sgn}(\mathrm{Re}[f(\varepsilon _{n}+\omega _{n})])}
{\sqrt{i\bar{f}(\bar{\varepsilon}_{n}+\omega _{n})+\bar{\mu}-\bar{P}(\bar{\varepsilon}_{n}+\omega _{n})}}
+
\right.
\\
\left.
\frac
{\mathrm{sgn}(\mathrm{Re}[f(\varepsilon_{n})])}
{\sqrt{i\bar{f}(\bar{\varepsilon}_{n})+\bar{\mu}-\bar{P}(\bar{\varepsilon}_n)}}
\frac
{\mathrm{sgn}(\mathrm{Re}[f(\varepsilon _{n}+\omega _{n})])}
{\sqrt{i\bar{f}(\bar{\varepsilon}_{n}+\omega _{n})+\bar{\mu}+\bar{P}(\bar{\varepsilon}_{n}+\omega _{n})}}
\right]
.
\end{gathered}
\label{sf_Pom}
\end{equation}
\end{centering}
\end{figure*}
Note that the polarization operator $\bar{\Omega}(\bar{\omega}_{n})-\bar{\omega}_{n}^{2}$ contains a factor $\sqrt{v_{s}^{2}/\alpha\Gamma}$ absent in the fermionic self-energy part. This factor will also arise if one calculates the vertex correction, as there one has to integrate a product of fermionic Green's functions like in the polarization operator. This allows us to use $\sqrt{v_{s}^{2}/\alpha\Gamma}$ as a small parameter to justify the Eliashberg-like approximation given by Fig. \ref{figFeyn}. We shall not neglect, however, the polarization operator, as it behaves linearly at low frequencies and might outpower the initial quadratic dispersion.

\begin{figure}[tbp]
\includegraphics[width=\linewidth]{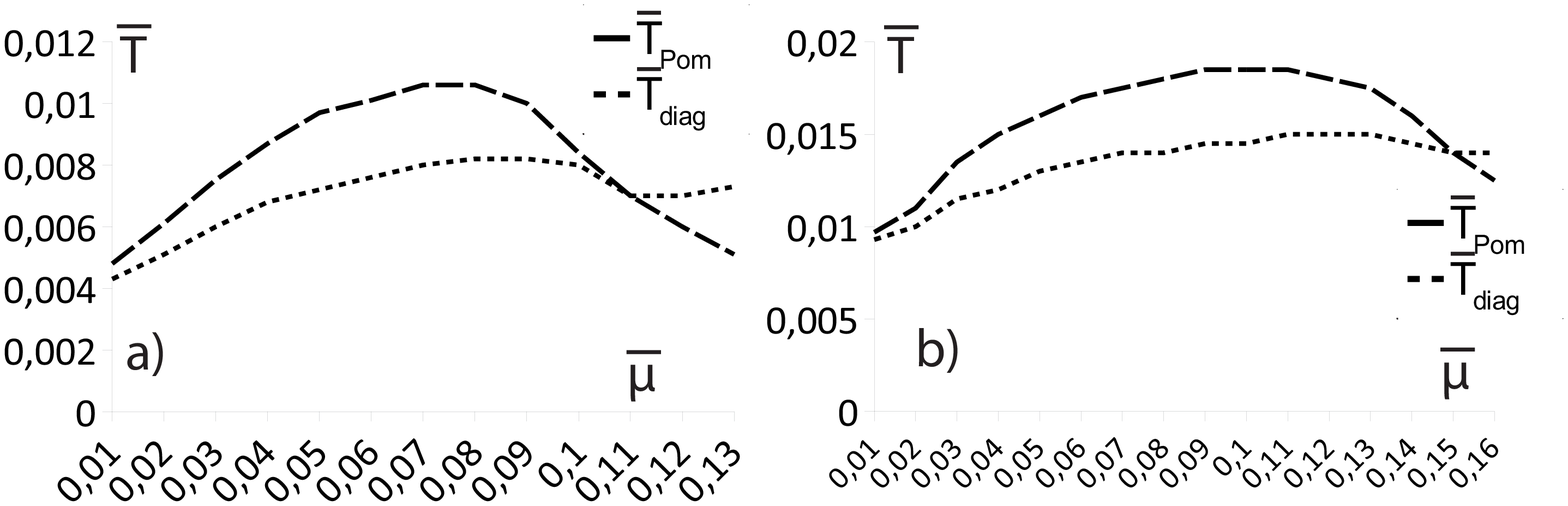}
\caption{$\bar{T}_{Pom}(\bar{\protect\mu})$ (dashed line) and $\bar{T}_{diag}(\bar{\protect%
\mu})$ (dotted line) for $\overline{(v_s/\xi)}=0.1$, $\sqrt{\frac{v_{s}^{2}/\alpha }{%
\Gamma }}=0.5(a),\;0.1(b)$.}
\label{figpomdiag}
\end{figure}

The equations (\ref{sf_Pom}) have been numerically solved by an iteration scheme, yielding the transition temperature $T_{Pom}$ where the 'odd' self-energy $P$ becomes non-zero. To show that this transition can be indeed leading we have also computed the transition temperature for a CDW with wavevector along the BZ diagonal. This instability has been found to be universally leading in previous studies. The transition temperature can be found from the linearized equation for the CDW order parameter $W_{diag}(\varepsilon_{n})$:
\begin{equation}
\begin{gathered}
\bar{W}_{diag}(\bar{\varepsilon}_{n})
=\frac{0.75\bar{T}}{2}\sum_{\varepsilon _{n}^{\prime
}}\frac{\bar{W}_{diag}(\bar{\varepsilon}_{n}^{\prime
})}{\sqrt{\bar{\Omega}(\bar{\varepsilon}_{n}-\bar{\varepsilon _{n}^{\prime
}})+\bar{a}}}
\\
\times \frac{\mathrm{sgn}\left(
\mathrm{Re}[f(\bar{\varepsilon}_{n}^{\prime })]\right)
}{\bar{f}(\bar{\varepsilon}_{n}^{\prime
})\sqrt{i\bar{f}(\bar{\varepsilon}_{n}^{\prime })+\bar{\mu}}}.
\end{gathered}
\label{sf_tdiag}
\end{equation}
The results of the numerical solutions are presented in Fig.\ref{figpomdiag}.

One can clearly see that for $\bar{\mu}$ less than a certain value Pomeranchuk instability is the leading one. Note that the ratio $\mu/T_{Pom}$ can be as high as $12$ for $\sqrt{v_{s}^{2}/\alpha\Gamma}=0.5$ and $9$ for $\sqrt{v_{s}^{2}/\alpha\Gamma}=0.1$.

As the Fermi Surface seen in ARPES experiments is universally found to be $C_4$-symmetric and in the light of the domained CDW structure \cite{Bi2212STM-3,y123REXS-4}, we assume that Pomeranchuk order should also be organized in domains with different sign of the order parameter. This constitutes a way of 'masking' a $C_4$ breaking alternative to the one proposed in \cite{yamase2009}.
\section{Incommensurate charge modulation.}
\begin{figure}[tbp]
\includegraphics[width=\linewidth]{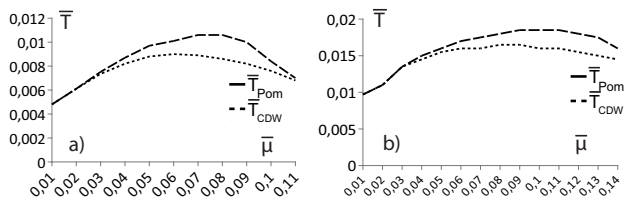}
\caption{$T_{Pom}(\bar{\protect\mu})$ (dashed line) and $\bar{T}_{CDW}(\bar{%
\protect\mu})$ (dotted line) determined from Eq. (\protect\ref{c11}) for $\overline{(v_s/\xi)}=0.1$, $\protect\sqrt{\frac{v_{s}^{2}/\protect\alpha }{\Gamma }}%
=0.5(a),\;0.1(b)$.}
\label{figCDW}
\end{figure}

The deformed Fermi surface of Fig.\ref{figpom} can be unstable to CDW formation at lower temperatures. The direction and the magnitude of the wavevector are directly related to the sign and the magnitude of $P$. We assume that the CDW wavevector should yield nesting in the region where the FS 'expands' due to the FS deformation. Then one has:
\begin{equation}
Q^{SF}(T)=2\sqrt{(\mu +0.5\left\vert P(-\pi T)+P(\pi T)\right\vert )/\alpha }.  \label{sf_q}
\end{equation}
In our model the FS in the second region 'closes' moving out of the considered region for $P>\mu$. However, as is seen from Fig.\ref{figpom}, in reality such a deformation can lead to emergent nesting in this region with the same vector direction as in the first one. The best-case scenario is that the nesting vectors in both regions coincide also in magnitude, {\it i.e.} $Q_1=Q_2$ (see Fig.\ref{figpom}). We shall assume that this is indeed the case, thus providing an upper limit on the $T_{CDW}$. In this case the equation for the CDW transition is:
\begin{eqnarray}
&&\bar{W}(\bar{\varepsilon}_{n})=0.75\;i\frac{\bar{T}_{CDW}}{2}%
\sum_{\varepsilon _{n}^{\prime }}\frac{\bar{W}(\bar{\varepsilon}_{n}^{\prime
})}{\sqrt{\bar{\Omega}(\bar{\varepsilon}_{n}-\bar{\varepsilon _{n}^{\prime }}%
)+\bar{a}}}  \notag \\
&&\times \frac{\mathrm{sgn}(\mathrm{Re}[f(\bar{\varepsilon}_{n}^{\prime })])%
}{(\left[ i\bar{f}(\bar{\varepsilon}_{n}^{\prime })+\bar{P}(\bar{\varepsilon}%
_{n}^{\prime })-P(0)\right] g\left( \bar{\varepsilon}^{\prime }\right) }.
\label{c11}
\end{eqnarray}%
The results of numerical calculations are presented in Fig.\ref{figCDW}. It turns out that the CDW transition can closely follow the onset of the FS deformation.

\section{Comparison with experiments and conclusions.}

\label{sec4}

Motivated by the existing ARPES data \cite%
{Bi2201ARPES-1,Bi2201ARPES-2,Bi2212ARPES-1} we have considered the SF model
with overlapping hotspots and demonstrated that the d-wave Fermi
surface distortion can be the leading instability. The
transition is further followed at a lower temperature by a transition into
a state with a d-form factor CDW directed along one of the BZ axes. The corresponding transition temperatures $T_{Pom}$ and $T_{CDW}$ can be not far away from each other.

The results obtained allow us to draw the following qualitative picture of
the charge order formation:

$\bullet $ At $T_{Pom}\geq T^{\ast }$ $C_{4}$ symmetry is broken by a Pomeranchuk
transition. The Fermi surface is deformed(see Fig. \ref{figpom}) and doped holes
are redistributed between the oxygen orbitals of the unit cell. The sample consists
of domains with different signs of the order parameter corresponding to two alternatives presented in Fig.\ref{figpom}.

$\bullet $ At $T_{CDW}<T_{Pom}$ a uniaxial d-form factor CDW forms in each domain.  CDW wave vector
is along one of the BZ axes dependinig on the sign of the Pomeranchuk order parameter inside each domain (see Fig.\ref{figpom}). The CDW period generally exceeds the one corresponding to antinodal nesting and is determined
self-consistently by the interaction and parameters of the Fermi surface. Qualitatively, the CDW wavevector tracks the FS and should decrease with hole doping (thus the CDW period should increase).

Our findings help us in understanding the results of recent experiments. The
Pomeranchuk deformation explains well why the C$_{4}$-symmetry at
commensurate peaks in Fourier transformed STM data \cite{Bi2212STM-0}. Formation of domains with different directions of the C$_{4}$-symmetry breaking is seen in STM experiments \cite{Bi2212STM-3} and can
also help explaining results of the transport measurements in YBCO \cite%
{y123Transp-2}. It is important to note, though, that the effects of the
deformation of the Fermi surface on transport can be masked by existence of
the domains. This may also resolve the apparent contradiction to the ARPES
data \cite{Bi2201ARPES-1,Bi2212ARPES-1} always showing a C$_{4}$-symmetric
Fermi surface.

The most important aspect of the Pomeranchuk order is that it explains the
robustness of the axial d-form factor CDW in the cuprates. We also note that
the organization of the CDW phase in the unidirectional domains is indeed
seen in STM \cite{Bi2212STM-3} and XRD \cite{y123REXS-4,HgXRD} measurements.
The coexistence of the unidirectional CDW and Pomeranchuk order also allows one to resolve
a seeming contradiction to results obtained in experiments on quantum
oscillations \cite{seb1}. Although the unidirectional CDW leads
to an open Fermi surface that does not support quantum oscillations, the
simultaneous presence of a C$_{4}$-symmetry breaking can indeed \cite%
{kivelson2011} close the Fermi surface leading to quantum oscillations in
high magnetic fields.

\begin{acknowledgements}
The authors gratefully acknowledge the financial support of the Ministry of
Education and Science of the Russian Federation in the framework of Increase
Competitiveness Program of NUST~\textquotedblleft MISiS\textquotedblright\
(Nr.~K2-2014-015).
\end{acknowledgements}


%
%

\end{document}